\documentstyle[11pt,newpasp,twoside,epsf]{article}
\def\edcomment#1{\iffalse\marginpar{\raggedright\sl#1\/}\else\relax\fi}
\reversemarginpar

\begin{document}
\title{From Infall to Rotation around Young Stars: The Origin of
Protoplanetary Disks}
\author{Michiel R. Hogerheijde}
\affil{Steward Observatory, The University of Arizona, 933 N. Cherry
 Ave, Tucson, AZ 85721-0065, USA; and Sterrewacht Leiden, Postbus
 9513, 2300 RA, Leiden, The Netherlands}

\begin{abstract}
The origin of disks surrounding young stars has direct implications
for our understanding of the formation of planetary systems. In the
interstellar clouds from which star form, angular momentum is
regulated by magnetic fields, preventing the spin up of contracting
cores. When $\sim 0.03$~pc-sized dense cores decouple from the
magnetic field and collapse dynamically, $\sim 10^{-3}$~km~s$^{-1}$~pc
of specific angular momentum is locked into the system. A viscous
accretion disk is one of two possible mechanisms available for the
necessary redistribution of angular momentum; the other one is the
formation of a multiple stellar system. Recent observational results
involving high-angular resolution observations are reviewed: the
presence of disks deep inside collapsing envelopes; an accretion shock
surrounding a disk; the velocity field in collapsing and slowly
rotating envelopes; a possible transitional object, characterized as a
large, contracting disk; and the velocity field in disks around
T~Tauri stars. Observational facilities becoming available over the
next several years promise to offer significant progress in the study
of the origin of protoplanetary disks.
\end{abstract}

\section{Introduction}

This contribution, the first in a series in this volume on the disks
that surround many young stars (e.g., Beckwith \& Sargent 1996),
reviews our current understanding of the transition from infalling
envelopes to rotating disks. It is based on recent theoretical and
observational work, with an emphasis on high-resolution results. Time
and space constraints limit the discussion to low-mass stars only, and
the intriguing question of the presence and nature of disks around
higher mass stars is left unaddressed (cf.\ Cesaroni et al.\ 1997;
Norris et al.\ 1998; Zhang, Hunter, \& Sridharan 1998; De Buizer 2003;
Sandell, Wright, \& Forster 2003).

The origin of a disk lies in the angular momentum of the cloud cores
from which the star condensed. The evolution of the angular momentum
during the star-formation process determines the properties of the
disk that will form: its mass and size, and therefore its
density. These disks are the likely birthplace of planets, and the
surface density in the disk is an important parameter in
planet-formation theories (e.g., Lissauer 1993; Ruden 1999). It is for
this reason that the study of the origin of disks is so important.

The outline of this contribution is as follows. Section~2 describes
the initial conditions: the angular-momentum distribution in dense
cloud cores before collapse. Section~3 presents evidence for a
characteristic length scale at which dense cores decouple from
the surrounding cloud and start to collapse. Section~4 gives a brief
overview of theoretical work on the collapse of rotating cloud
cores. Section~5 provides a sampling of recent high-resolution
observational work that addresses the origin of disks. Section~6
concludes this contribution with a short summary and an outlook to
future developments in this area.

\section{Initial Conditions: Rotation in Interstellar Clouds}

Interstellar cloud complexes have a hierarchical structure, consisting
of filaments, clumps, and dense cores (see, e.g., Blitz \& Williams
1999; McKee 1999; Myers 1999). It is generally thought that
magneto-hydrodynamic (MHD) turbulence provides the support of these
clouds against collapse under their own gravity, because the widths of
molecular emission lines are usually much larger than can be explained
by thermal motions alone (e.g., Fuller \& Myers 1992). The
quasi-equilibrium nature of clouds was recently questioned by
Hartmann, Ballesteros-Paredes, \& Bergin (2001), who argue that at
least the Taurus-Auriga cloud complex may have a much shorter life
time than previously thought. This would obviate the need for a
support mechanism, but disucssion about this interpretation is not yet
settled.

In addition to random (turbulent) motions, cloud cores also exhibit
organized motions (velocity gradients). Goodman et al.\ (1993)
constructed a velocity-gradient vs.\ size plot for a large sample of
dense cores, akin to the line-width vs.\ size relation first
identified by Larson (1981). Goodman et al.\ finds that the rotation
rate $\Omega$ increases at most moderately with decreasing core size,
$\Omega \propto R^{-0.4}$, although the relation shows significant
scatter. With assumptions about the mass distribution inside these
cores, this relation corresponds to a relation between the specific
angular momentum (=angular momentum per mass unit) $j\propto
R^{1.6}$. Both relations show that smaller cores do not spin much
faster, or, in other words, if smaller cores are the descendants
through contraction of larger cores, they do not spin up much and must
loose specific angular momentum. A likely explanation is that the
magnetic field lines that permeate the core and that are coupled to
the ions transfer angular momentum from the contracting to core to the
surrounding cloud. This mechanism is known as magnetic braking and was
studied in the context of weakly ionized media by K\"onigl
(1987). Another result of Goodman et al.\ (1993) is that the ratio of
rotational energy to gravitational energy of the cores, $\beta$, is
independent of size and $\approx 0.03$. Since this value is $\ll 1$,
it indicates that rotation is dynamically unimportant for cloud cores.

Recent theoretical work tries to reproduce the observed cloud
characteristics through numerical simulations of MHD turbulence (e.g.,
Myers \& Gammie 1999; Burkert \& Bodenheimer 2000; Ostriker et al.\
2001; Ossenkopf \& MacLow 2002; contribution by V\'azquez-Semadeni in
this volume). Burkert \& Bodenheimer (2000) find that the specific
angular momentum and the ratio of rotational to gravitational energy
on cloud-core scales in their simulations corresponds well to the
values reported by Goodman et al.\ (1993). This supports the idea that
MHD turbulence is the shaping mechanism of interstellar clouds and
regulates the angular momentum distribution of their constituents.

\section{Decoupling of Cloud Cores}

Magnetic fields cannot support cloud cores against collapse
indefinitely, because only the ions couple to the field lines while
the neutrals can slip through. Through this process, called ambipolar
diffusion, cloud condensations will slowly form. Because of the
increased recombination rate in denser gas, the ionization degree in
these cores will decrease, and dynamic collapse is inevitable. For
a more detailed description see, e.g., Shu in this volume.

Goodman et al.\ (1998) show evidence of a transition around 0.03~pc,
where the line widths (tracing random motions in the cores) become
much smaller (close to thermal) and no longer depend on the size of
the core. They interpret this as a cutoff scale for MHD
turbulence. When these MHD waves can no longer penetrate, the motions
inside the cores damp out rapidly.

Ohashi et al.\ (1997) find that around that same scale of 0.03~pc, the
specific angular momentum of objects no longer decreases with size, as
it did for cloud cores: infalling envelopes around Class 0/I objects
and rotationally supported disks all have angular momenta of $\sim
10^{-3}$~km~s$^{-1}$~pc independent of size. This amount of specific
angular momentum is comparable to the high end of values found in
Solar-neighborhood binaries (Heacox 1998). This further supports the
suspicion that when cores decouple from the magnetic field of the
surrounding cloud around 0.03~pc, $\sim 10^{-3}$~km~s$^{-1}$~pc of
specific angular momentum is locked into the nascent stellar
system.

\section{Theory of Collapsing and Rotating Cores}

Since rotation is dynamically unimportant in cores, it can initially
be treated as a perturbation to a non-rotating collapse solution. In
the absence of angular momentum redistribution, the infalling material
will spin up, and the perturbation approach is valid only outside the
radius where the centripetal force grows to a significant fraction of
the gravitational force (or, equivalently, where the stream lines of
the infalling material start to deviate appreciably from the radial
direction). Inside this rotational radius $R_c$ a disk forms (see
below). $R_c$ evolves over time as material with different amounts of
angular momentum falls in.

Terebey, Shu, \& Cassen (1984) based such a perturbational procedure
on the self-similar collapse solution of Shu (1977) and the assumption
of initial solid-body rotation. Because of this assumption of
solid-body rotation, the rotational radius grows rapidly with time,
$R_c \propto t^3$, as material that falls in at later times originated
from larger radii and carries more angular momentum. 

For a magnetized core a different initial angular momentum profile is
expected. Basu (1997) explores the slow contraction of a magnetized
core by ambipolar diffusion, and finds that the core tends toward
differential rotation. The outside of the core, where densities are
low and the ionization degree significant, remains strongly coupled to
the slow rotation of the surrounding cloud. The contracting central
region, less strongly coupled to the magnetic field, spins up, and is
joined by a sharp gradient in rotation rate to the outside. As a
result of this differentially rotating configuration, once dynamic
collapse starts the rotational radius grows only linearly with time,
$R_c \propto t$ (Basu 1998).

Whatever the rate of growth of $R_c$, within this radius material will
form a disk. Stahler et al.\ (1994) explored this analytically in the
Terebey et al.\ framework, and find that material piles up at ${1\over
3} R_c$. In their model an infinitely thin ring of infinitely high
density is formed; in reality, viscosity will spread this ring and set
up an accretion disk within this radius. Many authors have explored
the formation of a disk numerically (e.g., Bodenheimer et al.\ 1990;
Yorke, Bodenheimer, \& Laughlin 1993, 1995; Nakamura 2000;
Krasnopolsky \& K\"onigl 2002). Because of the difficulty of the
problem (the order of magnitude of scales involved; the physics of
magnetic fields; the poorly understood source(s) of viscosity) much of
this work is limited in the range of modeled scales or ages, and few
direct comparisons with observations have been made.

Other contributions in this volume discusses the structure of disks
around young stars in detail (see the contributions by Dullemond, by
Wardle, and by D'Alessio). Here it suffices to mention that disks
offer mechanisms to redistribute angular momentum for the first time
in the star-formation process since decoupling from the cloud magnetic
field. In addition to viscosity, disk instabilities can transfer
angular momentum, and photo-evaporated material can carry it
away. Alternatively, angular momentum can be stored in a binary
system, or carried away when objects are ejected from multiple systems.

\section{A Sampling of Recent Observations}

\subsection{Disks Deeply Embedded in Collapsing Cores}

Disks are formed inside collapsing cores, hampering direct observation
of their formation process. Aperture-synthesis observations at (sub)
millimeter wavelengths can reveal the presence of a distinct compact
source inside an extended envelope from the distribution of detected
flux as function of angular scale. Looney, Mundy, \& Welch (2000,
2003) surveyed a number of deeply embedded young stellar objects
(Class~0 YSOs) using the Berkeley-Illinois-Maryland Association (BIMA)
array. They find that during this early evolutionary stage the disks
are not more massive than during the later Class~I and T~Tauri stages,
with derived masses of $\le 0.12$~M$_\odot$. This suggests that the
disks funnel mass onto the star at a rate comparable to that with
which the envelope material falls onto the star--disk system.

Looney et al.\ also report that $>85$\% of the millimeter flux
originates from the extended envelope, illustrating the difficulty of
distinguishing a disk from the envelope. This is further exacerbated
by the fact that the envelopes are strongly centrally
concentrated. When trying to determine the degree of central
concentration from lower-resolution (e.g., SCUBA) data, the presence
of a disks affects the derived density slope when unaccounted for
(see, e.g, Harvey et al.\ 2003; J{\o}rgensen et al.\ submitted to
A\&A.). Fig.~1 illustrates how interferometer data can separate
envelope from disk emission.

\begin{figure}
%\plotone{hogerheijde_m_fig1.eps}
\plotfiddle{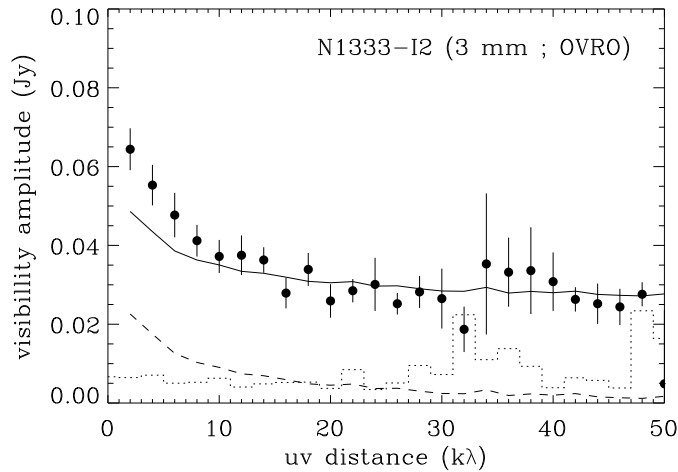}{5.5cm}{0}{55}{55}{-150}{-10}
\caption{Visibility amplitude vs.\ projected $uv$-distance
of the 3~mm emission from the embedded YSO NGC~1333~IRAS2, observed
with OVRO. The upturn of the flux at short $uv$ spacings reflects the
emission from the resolved envelope. The constant flux level of $\sim
22$~mJy at larger $uv$ spacings corresponds to the flux of the
unresolved disk. The dashed and solid curves show models without,
respectively, with a point source included. From J{\o}rgensen et al.\
(submitted to A\&A).}
\end{figure}

\subsection{An accretion shock}

When material falls in from the collapsing envelope onto the
rotationally supported disk, an accretion shock is expected. Recently,
Velusamy et al.\ (2002) presented evidence for such a
shock. Observations obtained with the Owens Valley (OVRO) Millimeter
Array of the Class~0 object L1157 show CH$_3$OH emission surrounding
the continuum disk, with respective sizes of $1210\times 640$~AU and
$800\times 590$~AU. Since CH$_3$OH is usually only found when icy
grain mantles evaporate, requiring temperatures higher than present in
the collapsing envelope, its presence is taken as evidence for
an accretion shock. This work illustrates how understanding the
gas- and solid-state chemistry is an essential tool.

\subsection{The velocity field in a Class~0 envelope}

Infall is commonly detected in Class~0 envelopes (Gregersen et al.\
1997; Mardones et al.\ 1997), while slow rotation is reported in
flattened Class~I envelopes (Ohashi, this volume). To date one of the
best studied velocity fields of a Class~0 object is that of
IRAM~04191+1522. Using molecular-line observations from the IRAM
30-meter telescope and the Plateau de Bure interferometer, Belloche et
al.\ (2002) find two regimes of inward motions: slow contraction at
$R>3000$~AU and inside-out collapse at $R<3000$~AU. They also identify
two regimes of rotation: a steep fall off in rotation rate at
$R>3500$~AU and a much shallower or even constant profile at
$R<3500$~AU. They suggest that this 3000--3500~AU (0.015--0.017~pc)
scale reflects the decoupling of the core from the magnetic field of
the surrounding cloud, as expected in models of the contraction and
collapse of magnetized cores (Basu 1997, 1998).

\subsection{L1489~IRS: A Transitional Object?}

One of the most intriguing observational pieces of evidence on the
transition from infall to rotation is offered by the object
L1489~IRS. Millimeter and sub-millimeter data presented by Hogerheijde
\& Sandell (2000) and Hogerheijde (2001) show that this Class~I object
is surrounded, not by an infalling envelope with perhaps a small
measure of rotation, but by a 2000~AU radius rotating
disk. Unresolved, single-dish spectra of the dense-gas tracers HCO$^+$
$J$=1--0, 3--2, and 4--3 show the `classic' infall signature (as
explained in Evans 1999). High-resolution interferometric images of
HCO$^+$ 1--0 and 3--2 emission obtained with OVRO and BIMA show a
highly flattened circumstellar configuration and a velocity gradient
clearly indicative of rotation (Fig.~2). The disk interpretation is further
supported by HST/NICMOS data from Padgett et al.\ (1999) that shows
light reflected off two flared surfaces, separated by a dark lane.
With a mass of 0.02~M$_\odot$, this disk is no different than commonly
found around T~Tauri stars (e.g., Beckwith \& Sargent 1996), but its
size (2000~AU radius) is larger by a factor of at least 2--3 than
other gas disks. This implies that its surface density is lower by an
order of magnitude or more: a fact that would have consequences for
its planet-forming capability.

\begin{figure}
%\plotone{hogerheijde_m_fig2.eps}
\plotfiddle{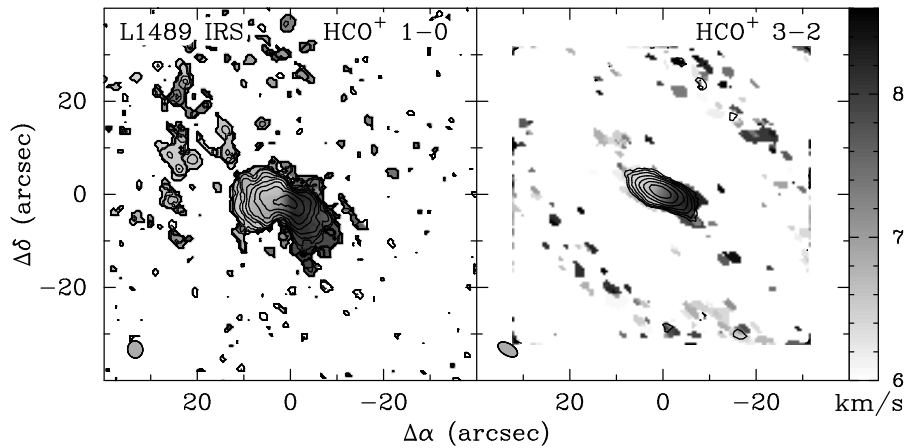}{5.5cm}{0}{65}{65}{-200}{-35}
\caption{Aperture-synthesis images of HCO$^+$ $J$=1--0 (left) and 3--2
(right) emission of L1489~IRS obtained at OVRO and BIMA, from
Hogerheijde (2001). The greyscale shows the velocity centroid of the
emission; the contours show integrated intensity. The HCO$^+$ 3--2
image is one of the highest frequency line images ever obtained at
BIMA.}
\end{figure}

Modeling the velocity field as observed in the interferometer beam in
detail, Hogerheijde (2001) finds that it is well characterized by
Keplerian rotation around a 0.65~M$_\odot$ object, but only if inward
motions amounting to 10\% of total magnitude of the velocity vector
are included. These radial motions are required to explain the
asymmetry of the HCO$^+$ position--velocity diagram and the clear
collapse signatures in the single-dish HCO$^+$ line profiles. With
such inward motions, the lifetime of this 2000~AU disk is only
$2\times 10^4$~yr. This velocity field is radically different from
those of Class~I objects, where rotation does not account for more
than a small fraction of the total velocity vector (Fig.~3).

\begin{figure}
%\plotone{hogerheijde_m_fig3.eps}
\plotfiddle{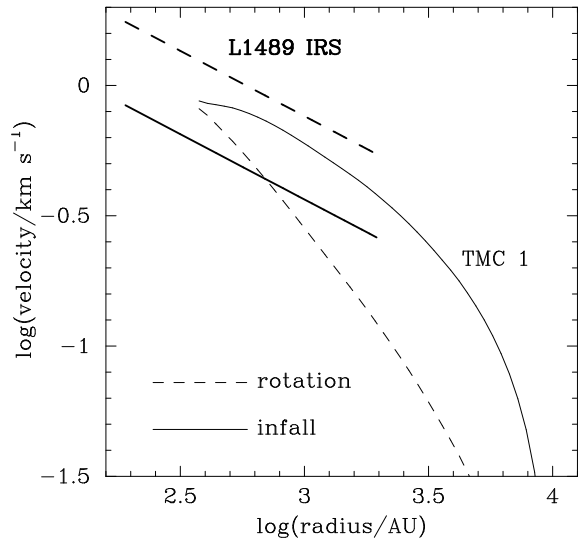}{6.5cm}{0}{40}{40}{-120}{0}
\caption{Comparison of the velocity fields around L1489~IRS and
TMC~1. The former is dominated by rotation with a $\sim 10$\%
contribution from infall; the latter is dominated by infall, with
rotation dynamically unimportant. From Hogerheijde (2001).}
\end{figure}

The resolution of the HCO$^+$ interferometer images is $\sim 5''$ or
700~AU. Do the inward motions continue all the way to the star? These
much smaller scales are sampled by the 4.65~$\mu$m Keck/NIRSPEC
spectrum presented by Boogert, Hogerheijde, \& Blake (2002) and
reproduced here in Fig.~4. This spectrum shows a wealth of gas-phase
CO absorption lines from the vibrational ground state into the first
excited state, in addition to a prominent solid-state CO absorption
band. The gas-phase lines from $^{12}$CO, $^{13}$CO, and C$^{18}$O
originate in a wide range of rotationally excited levels, requiring
kinetic temperatures as high as 250~K. Such temperatures are only
expected close to the star, well within an AU. The excellent spectral
resolution of NIRSPEC ($R\approx 25,000$) resolves the $^{12}$CO
absorption lines into a narrow line core plus a red wing extending to
100~km~s$^{-1}$ (Fig.~5). On the basis of the velocity model derived from the
HCO$^+$ interferometer data, such inward motions are expected only
within 0.1~AU. With a few minor adaptations (a change in the uncertain
density profile of the disk, and the inclusion of a small amount of
scattered light) this velocity model can provide an adequate fit to
the $^{12}$CO and $^{13}$CO absorption lines. Apparently, inward
motions continue from the 1000~AU scales sampled by the HCO$^+$ data
to within 0.1~AU as traced by the Keck M-band spectrum.

\begin{figure}
%\plotone{hogerheijde_m_fig4.eps}
\plotfiddle{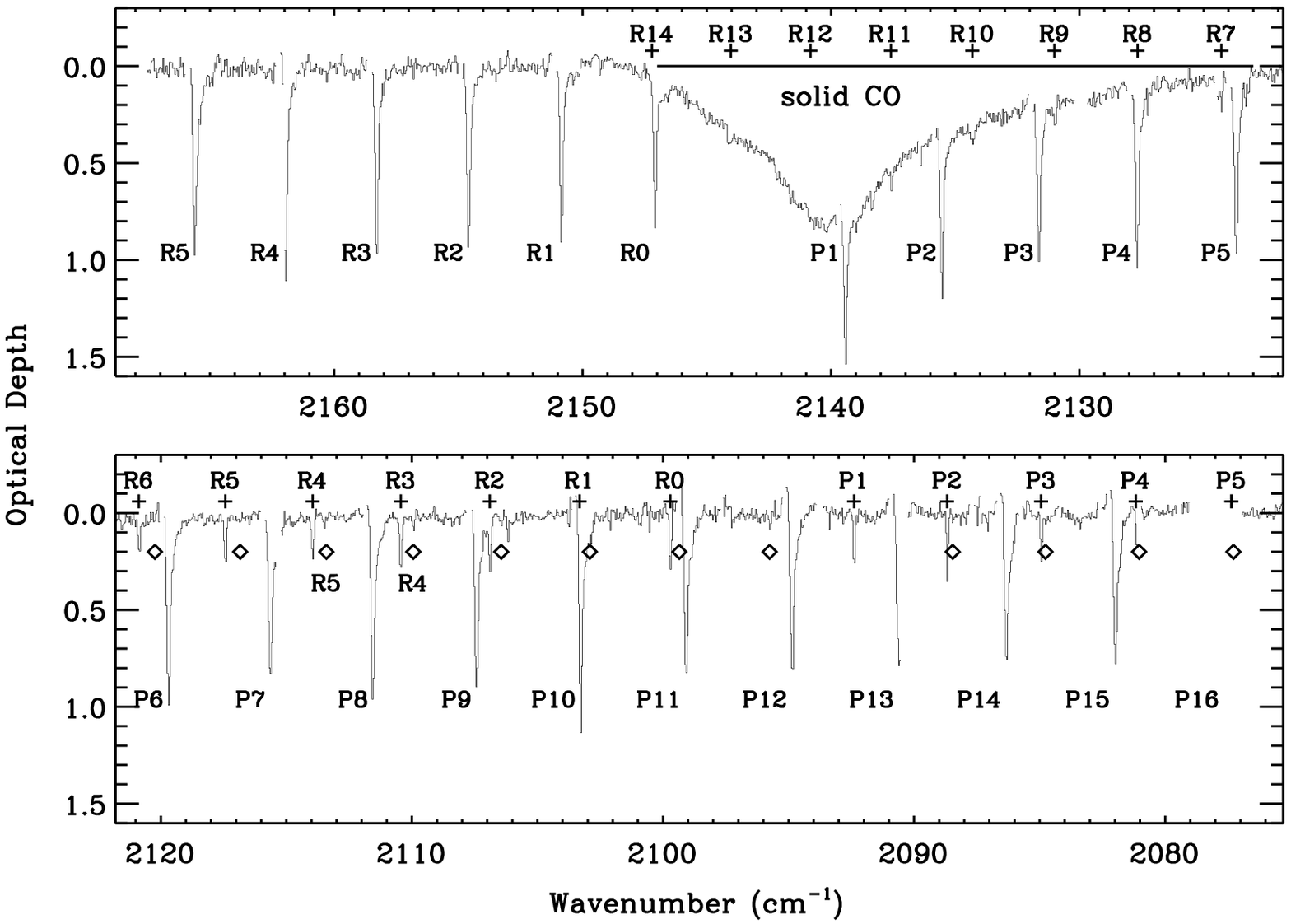}{6.5cm}{0}{65}{65}{-150}{-10}
\caption{Keck/NIRSPEC M-band spectrum of L1489~IRS showing narrow
absorption lines of $^{12}$CO, $^{13}$CO (marked by $+$), and
C$^{18}$O (marked by $\diamond$), as well as a broad absorption
feature of solid CO ice. From Boogert et al.\ (2002).}
\end{figure}

\begin{figure}
%\plotone{hogerheijde_m_fig5.eps}
\plotfiddle{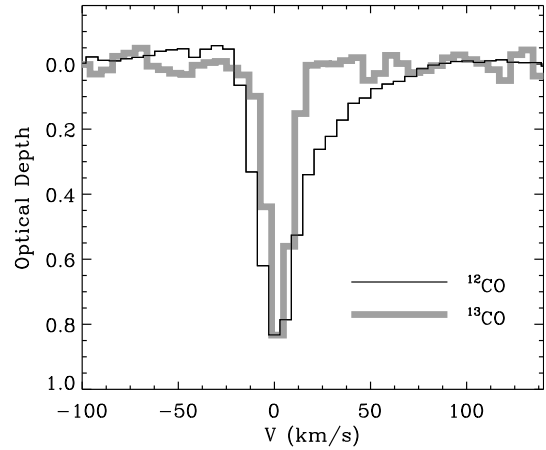}{5.5cm}{0}{50}{50}{-100}{0}
\caption{Comparison of the $^{12}$CO (thin black line) and $^{13}$CO
(thick grey line) spectra of L1489~IRS, averaged over the $^{12}$CO
P(6)--P(15) and all $^{13}$CO lines. The $^{13}$CO lines are scaled to
the same depth as the $^{12}$CO lines, to bring out the different line
shapes. From Boogert et al.\ (2002).}
\end{figure}

This velocity field does pose some problems, however. If the entire
disk participates in the inward motions, the implied accretion rate
exceeds the observational limits (Muzerolle, Hartmann, \& Calvet 1996)
by more than an order of magnitude. Also, a viscous accretion disk
model would predict subsonic flow, while the observed velocities are
supersonic. A possible solution would be inflow in a thin surface
layer only (reducing the amount of mass participating in the flow and
hence the accretion rate). Another solution would be that the
observations trace two regions if inflow: one in the outer disk ($\ga
500$~AU) probed by the HCO$^+$ interferometry where envelope material
is settling onto the (unresolved) rotationally supported disk; and a
second in the inner disk ($\la 0.1$~AU) where material accretes from
the disk onto the star. The fact that the same velocity model fits
both regions is not surprising, since it is determined by the
gravitational potential well of the same star. Both explanations are
supported by the failure of the velocity model to fit the core of the
$^{12}$CO absorption line, underestimating the amount of the material
that is not moving radially. The alternative explanation of the
$^{12}$CO lines originating from two disks around the members of a
binary system is ruled out by the NICMOS observations of Padgett et
al.\ (1999) and lack of temporal changes in the line profiles
(Boogert, priv.\ comm.).

In summary, L1489~IRS appears to represent a transitional object
between Class~I YSOs characterized by infalling envelopes with little
rotation, and T~Tauri stars surrounded by Keplerian disks. Inward
motions are present throughout the large, rotating disk that
surrounds L1489~IRS, but their exact nature remains elusive. Higher
(both angular and spectral) resolution observations and the
identification of additional object like it are required to correctly
interpret the nature of this intriguing object.

\subsection{Keplerian Disks around T~Tauri Stars}

Other contributions in this volume (e.g., Wilner, Dutrey) discuss
observations of disks around T~Tauri stars in detail. Here it only
needs to be mentioned that their velocity fields are Keplerian (e.g.,
Simon, Dutrey, \& Guilloteau 2000) with little indication for large
radial motions. In viscous accretion disks, no such motions are
expected. The presence of accretion signatures in the stellar spectra
clearly shows that material does flow inward through these disks,
however.

\section{Summary and Outlook}

Before collapse, the angular momentum in cloud cores is regulated by
MHD turbulence. Through ambipolar diffusion, dense cores slowly grow,
which have a decreasing ionization degree and which ultimately
decouple from the magnetic field and collapse. The scale for this
decoupling appears to be $\sim 0.03$~pc. At the moment of decoupling,
angular momentum is locked in the collapsing core, which will lead to
the formation of a disk. Theoretical description of the disk-formation
process are progressing, but detailed comparisons with observations
are still lacking. Observationally, millimeter-interferometry can pick
out disks deep inside envelopes. These disks are found to be no more
massive than their counterparts during later evolutionary stages,
indicating they funnel mass onto the star efficiently. The velocity
fields of only a handful of YSOs have been studied in detail, but they
appear to follow theoretical expectations for collapsing, magnetized
cores. The object L1489~IRS may represent a short-lived, transitional
object between these embedded YSOs and T~Tauri stars. It is surrounded
by a large (2000~AU radius), rotating disk, that has significant
inward motions traced from $\ga 700$~AU to $\la 0.1$~AU with HCO$^+$
interferometry and CO M-band spectroscopy. While the exact nature of
this object's velocity field remains unknown, L1489~IRS, and other
possible objects like it, offer intriguing insight into the transition
from infall to rotation.

The outlook for high-resolution observations is very promising, with
numerous facilities coming online, being constructed, or being
planned. (Sub) millimeter arrays such as the Combined Array for
Research in Millimeter Astronomy (CARMA), the SubMillimeter Array
(SMA), and ultimately the Atacama Large Millimeter Array (ALMA) will
revolutionize our view of collapsing envelopes and circumstellar
disks. The sensitivity and accuracy with which ALMA is designed to
probe a wide range of spatial scales will prove essential in
uncovering the birth of disks deep inside the collapsing
envelopes. Our increased understanding of the chemistry that occurs
during star formation will help us the select tracers that are as
unambiguous as possible, even for unresolved observations.

At shorter wavelengths, the spectral resolution at infrared
wavelengths is increasing incrementally, with the EXES instrument
designed for the Stratospheric Observatory For Infrared Astronomy
(SOFIA) surpassing the NIRSPEC data presented above by a factor of
almost 4 (but with lower sensitivity). At these infrared wavelengths,
interferometric facilities such as VLTI, Keck-I, and one day
Darwin/TPF will uncover the processes occurring close to the star,
such as accretion, outflow, and disk structure. Together, these
observations will no doubt inspire more theoretical work, and direct
comparisons between theory and observations. The prospects for
uncovering the origin of disks, and thereby the initial conditions of
planet formation, are excellent.

\acknowledgments

MRH wishes to thank the organizers for their invitation to speak at
the symposium, and regrets that unforeseen circumstances prevented him
from attending. He extends special thanks to Geoffrey Blake for
replacing him at short notice.

\end{document}